\documentclass[12pt]{amsart}
\usepackage{amsfonts,amssymb,amsmath}
\usepackage{url}
\usepackage{enumerate}
\usepackage{comment} 
\usepackage{color}
\usepackage[singlelinecheck=false]{caption}

\usepackage[left=2.3cm,right=2.5cm,top=2.5cm,bottom=2.2cm]{geometry}

\newtheorem{theorem}{Theorem}[section]
\newtheorem{lemma}[theorem]{Lemma}
\newtheorem{proposition}[theorem]{Proposition}
\newtheorem{corollary}[theorem]{Corollary}
\theoremstyle{definition}

\newtheorem{remark}[theorem]{Remark}
\numberwithin{equation}{section}

\newcommand{\esssup}{\textrm{-}{\rm ess.sup\;}}
\newcommand{\essinf}{\textrm{-}{\rm ess.inf\;}}

\newcommand{\eps}{\epsilon}

\newcommand{\C}{\mathbb{C}}

\newcommand{\SL}{\mathcal{H}}
\newcommand{\R}{\mathbb{R}}

\newcommand{\E}{\mathbb{E}}

\author{Vanderl\'ea R. Bazao}   
\address{Faculdade de Ci\^encias Exatas e Tecnologias, UFGD, Dourados, MS, 79804-970  Brazil}

\author{T\'ulio O. Carvalho}
\address{Departamento de Matem\'atica, UEL, CP 10011, Londrina, PR, 86057-970 Brazil}

\author{C\'esar R. de Oliveira}  
\address{Departamento de Matem\'atica, UFSCar, S\~ao Carlos, SP, 13560-970 Brazil}

\begin{document}

\title[Interval Hausdorff spectral measures]{Spectral Hausdorff dimensions for a class of Schr\"odinger operators in bounded intervals}

\begin{abstract}Exact Hausdorff dimensions are computed for  singular continuous components of the spectral measures of a class of Schr\"odinger operators in bounded intervals. 
\end{abstract}

\maketitle

\

%   SSSSS   
\section{Main results}\label{intro}

We are interested in Hausdorff dimensional properties of spectral measures of Schr\"odinger operators   
\begin{equation}\label{eq1}
    (Hu)(x)=-\frac{\mathrm{d}^2 u}{\mathrm{d}x^2}(x)+V(x)u(x)
\end{equation}
acting in $\mathrm{L}^2 (I_b)$, where $I_b=[0,b]$, $0<b<\infty$, is a bounded   interval of~$\R$; our potentials $V(x)$  are  signed combs of delta distributions carefully spaced in~$I_b$ and accumulating only at~$b$. The boundary condition at~$0$ is 
\begin{equation}\label{C.C.}
u(0) \cos(\varphi) +u'(0)\sin(\varphi)=0,
\end{equation}
with  $\varphi\in [0,  2\pi)$ fixed.

In~\cite{P1975}, Pearson has presented  a family  of such Schr\"odinger operators $-\mathrm d^{2}/\mathrm d x^{2}+V_0(x)$, on the bounded interval~$I_b$, which has  purely absolutely continuous spectrum in a certain range of energies (the potential $V_0(x)$ is a selected comb of delta distributions and its construction is recalled in Appendix~\ref{M.S.C.}). Borrowing ideas from~\cite{KLS2, P1978}, we will perturb such potential $V_0(x)$ to obtain a model~\eqref{eq1} with a  singular continuous spectral component in this interval (see also Example~$14.6.j$ in~\cite{P}). 

The main contribution of this work is to  compute the Hausdorff dimensions of their spectral measures. It is usually hard to present examples of potentials for which one can say something about fractal properties of spectral measures. To the best of our knowledge, these are the first examples,  in bounded intervals, of singular continuous Schr\"odinger operators with computable dimensions.   

We need some preparation in order to state our main results.

\subsection{Hausdorff dimension of measures}\label{subsectFDM}

If $\mu$ is a positive and finite Borel measure on~$\mathbb R$, its  \emph{lower local dimension} at~$x\in\mathrm{support}(\mu)$ is given by
\[
d_\mu^-(x) = \liminf_{\epsilon\downarrow 0}\frac{\ln\mu((x-\epsilon,x+\epsilon))}{\ln\epsilon}.
\] The general idea is to estimate the scaling property $\mu((x-\epsilon,x+\epsilon))\sim \epsilon^{d_\mu(x)}$ for small~$\epsilon>0$.

Let $ \dim_{\mathrm H}(S)$  denote the  Hausdorff  dimension of the set~$S\subset\mathbb R$ and $0< \alpha \le 1$; the {\em upper Hausdorff dimension} of~$\mu$ is defined as 
\[
\dim_{\mathrm H}^+(\mu)= \inf \{ \dim_{\mathrm H}(S); \mu(\mathbb R\setminus S)=0, \,S\; \mathrm{a\; Borel\; subset\; of}\; \mathbb R \},
\]
and its {\em lower Hausdorff dimension}   as
\[
\dim_{\mathrm H}^-(\mu)= \sup\{\alpha; \mu(S)=0\;\;\mathrm{if}\;\dim_{\mathrm H}(S)<\alpha,\;S\; \mathrm{a\; Borel\; subset\; of}\; \mathbb R\}.
\]
If $A \subset\mathbb{R}$ is a Borel set, we shall also consider the dimensions of the restriction of~$\mu$ to such set, that is, $\mu_{;A}(\cdot):= \mu(A\cap\cdot)$.

It turns out that~\cite{F,M} 
\begin{equation}
\dim_{\mathrm H}^-(\mu) = \mu\essinf d_\mu^- ,\qquad \dim_{\mathrm H}^+(\mu) = \mu\esssup d_\mu^-\,. \label{eqPdimLocalDim}
\end{equation}
Hence, the information of the lower local dimensions of~$\mu$ gives the values of its Hausdorff dimensions.

\subsection{The potential}\label{subsectPot}

Let the potential $V_0(x)$ be a  comb of delta distributions at suitable points $a_n\in (0,b)$, as constructed in~\cite{P1975,P1978,P},  for which the Schr\"odinger operator~\eqref{eq1} has purely absolutely continuous spectrum on an interval $J\subset \R$.  We consider the potential
\begin{equation}\label{psc}
V_\omega(x)=V_0(x)+\sum_{n=1}^{\infty}g_{n}^\omega\delta(x-b_n), \quad 0<x<b,
\end{equation}
with  $b_{n}=\sum_{j=1}^{n}8a_j$ and 
$b=\lim_{n\rightarrow\infty}b_{n}=\sum_{j=1}^{\infty}8a_j<\infty$; the 
potential~$V(x)$ is a perturbation of $V_0(x)$ by a~$\delta$ comb located at the 
points of the sequence~$ (b_ {n})$. 

We assume that $(g_n^{\omega})$ is a sequence of independent (real) random 
variables defined in a probability space  with (probability) 
measure~$\nu(\omega)$, and let~$\E$ denote the expectation with respect 
to~$\nu$. For each realization $\omega$, denote the corresponding Schr\"odinger operator (with boundary condition~\eqref{C.C.}) by
\begin{equation}\label{eqHomega}
H_{\omega}=-\frac{\mathrm d^2}{\mathrm d x^{2}}+V_\omega(x), \quad 0<x<b,
\end{equation}
and its spectral measure by~$\rho_\omega$.
We will also assume that there is $\lambda>0$ such that
\begin{enumerate}
	\item [(i)] $\E((g_{n}^{\omega})^{2})=\lambda^{2} n^{-1}$ and, for a 
positive constant~$C_4$,  $\E((g_n^\omega)^4) \le C_4\lambda^4n^{-2}$;
	\item [(ii)] $\E(g_{n}^{\omega})=0$;
	\item [(iii)] for some $\epsilon>0$, there is a  positive constant~$C_1$ so that $\sup_{\omega}|g_{n}^{\omega}|\leq C_1n^{-\frac13-\epsilon}$;
	\item [(iv)] $g_{n}^{\omega}$ is independent of 
$(g_{j}^{\omega})_{j=1}^{n-1}$.
	\end{enumerate}
Note that the hypothesis that the fourth moment scales like the square of the second moment is true for random variables following the normal distribution.  

\subsection{Main results}\label{subsecMainRes}

For $0<\lambda<2$, denote
\[J=J(\lambda):=\left(\frac{6-3\sqrt{4-\lambda^{2}}}{5},\frac{6+3\sqrt{4-\lambda^{2}}}{5}\right)\left\backslash \left\{\frac{3(2-\sqrt{2})}{5},\frac{6}{5}, \frac{3(2+\sqrt{2})}{5}\right\}\right.,
\]  and observe that $J(\lambda)\subset (0,12/5)$ for all~$0<\lambda<2$.

\begin{proposition}\label{thmSCspectrum}
Fix $\lambda\in (0,2)$. Then, for $\nu$-a.e.~$\omega$, $J(\lambda)$ is a subset of the spectrum of~$H_\omega$ and the restricted operator $H_\omega P^{H_\omega}(J(\lambda))$  is purely singular continuous (where $P^{H_\omega}$ is the spectral projection of~$H_\omega$).
\end{proposition}

To some extent, this proposition is similar to some of the results  discussed in Pearson \cite{P1975,P1978,P}, but here we have to deal with the role of random potentials, as considered in~\cite{KLS2}.

\begin{theorem}\label{thmExactDim} Fix $\lambda\in (0,2)$. Then, for $\nu$-a.e.~$\omega$ and each $E\in J(\lambda)$, the spectral measure~$\rho_\omega$ has lower local dimension
\begin{equation}\label{eqDefAlpha}
d_{\rho_\omega}^-(E)=\alpha=\alpha(E,\lambda):=1-\frac{9\lambda^{2}}{60E-25E^{2}}\,.
\end{equation}
\end{theorem}

\begin{corollary}\label{corolExactDim}
Let $\nu,\lambda,J,H_\omega$ and $\rho_\omega$ be as above. Then, for $\nu$-a.e.~$\omega$,
\begin{itemize}
\item[(i)] $\dim_{\mathrm H}^-(\rho_{\omega;J})=0 $ and $\dim_{\mathrm H}^+(\rho_{\omega;J})=1-\frac{\lambda^2}{4}$.
\item[(ii)] given an interval $[m,M]\subset (0,1-\lambda^2/4)$, $m<M$, there is a subset $J_{m,M}\subset J$ so that, for the spectral measure $\rho_\omega^{m,M}:=\rho_{\omega;J_{m,M}}$ of the restricted operator $H_\omega^{m,M}:=H_\omega\,P^{H_\omega}(J_{m,M})$, one has
\[
\dim_{\mathrm H}^-(\rho_\omega^{m,M})=m \quad \mathrm{and}\quad \dim_{\mathrm H}^+(\rho_\omega^{m,M})=M.
\]
\end{itemize}
\end{corollary}

In Section~\ref{sectProofPrinc}, we present the proofs of Proposition~\ref{thmSCspectrum}, Theorem~\ref{thmExactDim} and Corollary~\ref{corolExactDim}, which make use of nontrivial technical estimates (in particular Theorem~\ref{thmSubord} on $\alpha(E,\lambda)$-subordinate solutions) of the asymptotic behaviour of solutions to the eigenvalue equation for~$H_\omega$. Section~\ref{SeI} provides  results about Hausdorff  subordinacy in bounded intervals. In Section~\ref{solution} some of the techniques mentioned above are discussed, in order to prove  Theorem~\ref{thmSubord} in Section~\ref{sectProofThmSub}. For the reader's convenience,  some details regarding the construction of the unperturbed potential~$V_0$ are recalled in Appendix~\ref{M.S.C.}.

%   SSSSS
\section{Proofs of the main results}\label{sectProofPrinc}

\begin{proof}(Proposition~\ref{thmSCspectrum})
As discussed in Subsection~\ref{subsectDemoSingCont}, if  $\sum_{n=1}^\infty (g_n^\omega)^2=\infty$, then $J(\lambda)$ is contained in the spectrum of~$H_\omega$ and this operator is purely singular continuous there. Hence, the proof here amounts to show that~$\nu$-a.e.~$\omega$ one has $\sum_{n=1}^\infty (g_n^\omega)^2=\infty$.

Let   
$X_{k,\omega}=\sum_{n=1}^k (g_n^\omega)^2$ and, given $N>0$, set
\[ U_N=\Big\{ \omega\ ;\ \sum_{n=1}^\infty (g_n^\omega)^2 <N\Big\} \ . \]
Then $\nu(U_N)\leq \nu(X_{k,\omega}<N)$ for every $k$. Pick $k_0$ such that 
$\E(X_{k,\omega})>N/2$, for every $k\geq k_0$. 

From the classical Bienaym\'e-Chebyshev inequality, for every $t>0$, 
\[ 
\nu(\{\omega\ ;\ |X_{k,\omega}-\E(X_{k,\omega})|> t\}) \leq \frac{{\rm 
var}(X_{k,\omega})}{t^2}, 
\]
where ${\rm var}(\cdot)$ denotes the variance. By the independence of $g_n^\omega$, 
\[ {\rm var}(X_{k,\omega}) = \sum_{n=1}^k {\rm var}((g_n^\omega)^2) <  \tilde 
C_4  , 
\]
since $\E((g_n^\omega)^4) \le C_4\lambda^4n^{-2}$. 

Therefore, for each $k$, 
\[ 
\nu(\{\omega\ ;\ \E(X_{k,\omega})-X_{k,\omega}>t\}) \leq \nu(\{\omega\ ;\ 
|\E(X_{k,\omega})-X_{k,\omega}|>t\}) \leq \frac{\tilde 
C_4}{t^2} ,
\]
which implies that $\nu(\omega; X_{k,\omega}<N) < \frac{4\tilde C_4}{N^2}$ for $k\geq 
k_0$. 
Since $N$ is arbitrary, the result follows.
\end{proof}

For  a solution~$u$ to the eigenvalue equation
\begin{equation}\label{EAC}
    (H_\omega u)(x)=Eu(x)
\end{equation}
and $0<L<b$, denote 
\begin{equation}\label{eqL2ateL}
\|u\|_{L}=\left(\int_{0}^{L}|u(r)|^2 \mathrm d r\right)^{1/2}.
\end{equation} For a finite Borel measure~$\mu$ on~$\mathbb R$ and $0\le\gamma\le1$, consider the  {\em upper $\gamma$-derivative} of~$\mu$ at~$x\in\R$, given by
\[
 \overline D_\mu^\gamma(x) = \limsup_{\epsilon\to0} \frac{\mu((x-\eps,x+\eps))}{(2\epsilon)^\gamma}\,.
\]
The proof of Theorem~\ref{thmExactDim} is based on the following theorem, whose proof is presented in Section~\ref{sectProofThmSub} and it is an important technical part of this paper. Since for $\nu$-a.e.~$\omega$ the operator~$H_\omega$ has no eigenvalue in~$J$, for all  nonzero solutions~$u$ to the eigenvalue equation~\eqref{EAC} with~$E\in J$, one has $\|u\|_L\to\infty$ as $L\to b$.

\begin{theorem}\label{thmSubord}
Let $0<\lambda<2$,  and $\alpha(E,\lambda)$ be as on the right hand side of~\eqref{eqDefAlpha}. Then, for $\nu$-a.e.~$\omega$ and each $E\in J(\lambda)$, there exists a solution $u_E^S$ to~\eqref{EAC}  so that, for all other solutions~$u_E$, linearly independent with~$u_E^S$, 
\begin{equation}\label{eqAlfaSub}
\lim_{L\to b}\frac{\|u_E^S\|_{L}}{\|u_E\|_{L}^{\alpha/(2-\alpha)}}=A
\end{equation}
holds for some appropriate value of~$A\in(0,\infty)$.
\end{theorem}

The proof of Theorem~\ref{thmSubord} makes use of adaptations to the bounded interval setting of sparse potentials in unbounded intervals~\cite{KLS2}.  In a bounded interval, one does not have room for sparse potentials, which here will be replaced by signed delta comb potentials, with diverging intensities as one approaches the interval endpoint~$b$.

\begin{proof}(Theorem~\ref{thmExactDim})
First note that $y\mapsto y/(2-y)$ is a monotonically increasing function of 
$y\in[0,1]$. For $E\in J$ and $\alpha'<\alpha$, by~\eqref{eqAlfaSub} one has \[
\lim_{L\to b}\frac{\|u_E^S\|_{L}}{\|u_E\|_{L}^{\alpha'/(2-\alpha')}}=\infty,
\] and so, by the Hausdorff subordinacy theory (see Section~\ref{SeI} 
and~\cite{JL1}), it follows that $\overline D_{\rho_\omega}^{\alpha'}(E)=0.$ Hence, given $\delta>0$, for all 
$\epsilon>0$ small enough, 
\[
\frac{\rho_\omega((E-\eps,E+\eps))}{\eps^{\alpha'}}<\delta\; \Longrightarrow \; \frac{\ln \rho_\omega((E-\eps,E+\eps))}{\ln \eps}> \frac{\ln \delta}{\ln \eps}+\alpha',
\] and so $d_{\rho_\omega}^-(E)\ge \alpha'$; since this holds for all $\alpha'<\alpha$, one has $d_{\rho_\omega}^-(E)\ge \alpha$.

On the other hand, if $\alpha<\alpha'$, by~\eqref{eqAlfaSub} one has
\[
\lim_{L\to b}\frac{\|u_E^S\|_{L}}{\|u_E\|_{L}^{\alpha'/(2-\alpha')}}=0,
\] and so, again by the Hausdorff subordinacy theory, it follows that $\overline D_{\rho_\omega}^{\alpha'}(E)=\infty.$ Hence, given $N>0$, there is a subsequence $\epsilon_j\downarrow 0$ with 
\[
\frac{\rho_\omega((E-\eps_j,E+\eps_j))}{\eps_j^{\alpha'}}>N \; \Longrightarrow \; \frac{\ln \rho_\omega((E-\eps_j,E+\eps_j))}{\ln \eps_j}< \frac{\ln N}{\ln\eps_j}+\alpha'
\] and so $d_{\rho_\omega}^-(E)\le \alpha'$; since this holds for all $\alpha<\alpha'$, one has $d_{\rho_\omega}^-(E)\le \alpha$. By combining both inequalities, $d_{\rho_\omega}^-(E)= \alpha$.
\end{proof}

\begin{proof}(Corollary~\ref{corolExactDim})
(i)~By \eqref{eqPdimLocalDim} and Theorem~\ref{thmExactDim}, and since $\alpha(E,\lambda)$ is a continuous function of the variable~$E\in J$, it is enough to note that $\inf_{E\in J}\alpha(E,\lambda)=0$ and $\max_{E\in J}\alpha(E,\lambda)=1-\lambda^2/4.$

(ii)~It is enough to use the inverse function $\alpha^{-1}$ to pick 
$J_{m,M}=\alpha^{-1}(\cdot,\lambda)\big([m,M]\cap J(\lambda)\big)$.
\end{proof}

%  SSSSS
\section{Hausdorff subordinacy in bounded intervals} 
\label{SeI}
We provide in this section the necessary results about 
Hausdorff  subordinacy for operators~\eqref{eq1} acting in $\mathrm{L}^2 (I_b)$. 
We  depart from known results on subordinacy  for Schr\"odinger operators  with action~\eqref{eq1}  in the cases of 
unbounded intervals $I = \R$ or $I = [0, \infty)$~\cite{CO, JL1, JL2, 
JZ,DKL,KLS2}. The Hausdorff subordinacy results are generalizations of the 
subordinacy theory of Gilbert and Pearson~\cite{G, GP}. 

 The subordinacy theory for operators~\eqref{eq1}  in bounded intervals was discussed in~\cite{G,G2,P}. Now we dwell on the adaptation of such results to Hausdorff  subordinacy. We begin  by  discussing the spectral properties of operators 
\begin{equation}\label{SL}
\SL=-\frac{\mathrm d^2}{\mathrm dx^2} + V(x), 
\end{equation}
acting in $\mathrm L^2(I_b)$, that are regular at  the point~$0$ and limit point at~$b$ (see~\cite{P} for such standard concepts). A self-adjoint operator~$H$ can be obtained from $\SL$ in a suitable domain, satisfying the boundary condition~\eqref{C.C.} at point 0, with fixed $\varphi\in [0, \pi)$. By varying the boundary conditions, we obtain  a family of  self-adjoint operators $H$ resulting from~$\SL$.

The spectral function $\rho(E)$ of~$H$ generates a Borel-Stieltjes measure~$\rho$ which is the spectral measure associated with the operator $H$, which, by its turn, is related to the Weyl-Titchmarsh  $m$-function through
\[
\rho(E_2)-\rho(E_1)=\lim_{\delta\rightarrow 0}\lim_{\varepsilon\rightarrow 0}\frac{1}{\pi}\int_{E_{1}+\delta}^{E_{2}+\delta}\mathrm{Im}(m(E+i\varepsilon))\mathrm dE,
\]
and the inverse relation
\[
m(z)=\int_{-\infty}^{\infty}\frac{1}{(E-z)}\mathrm d\rho(E) +\cot(\varphi).
\]

The $m$-function $m(z)$ satisfies
\begin{equation}\label{mfuncaoC}
\hat{u}(x,z):=u_{2}(x,z)+m(z)u_{1}(x,z) \in {\mathrm L}^2(I_b),
\end{equation}
for all $z\in \C\backslash \R$, with $u_1(x,z)$ and $u_2(x,z)$ solutions to $\SL u=zu$ satisfying the orthogonal initial conditions
\begin{equation} \label{C.I.} \left\{ \begin{array}{ll}
u_{1}(0,z)=-\sin\varphi &\ \ \  u_{2}(0,z)=\cos\varphi
\\ u'_{1}(0,z)=\cos\varphi &\ \ \ u'_{2}(0,z)=\sin\varphi
\end{array} \right. .
\end{equation}

We denote by $u_{1,E}$ and $u_{2,E}$ the solutions to the equation~\eqref{EAC} in~$I_b$, satisfying the initial conditions~\eqref{C.I.} for $\varphi\in[0,2\pi)$ fixed. Since at least one of these solutions has unbounded norm, let $L(\varepsilon)\in (0,b)$ be the length defined by the equality (see~\eqref{eqL2ateL})
\begin{equation} \label{comp.l}
 \|u_{1,E}\|_{L(\varepsilon)} 
 \|u_{2,E}\|_{L(\varepsilon)}=
\frac{1}{2\varepsilon}\ . 
\end{equation}

Following the lines of Jitomirskaya-Last~\cite{JL1}, and taking into account the parameter variation formula, i.e., equation~$(7.2.8)$ in~\cite{P}, one can prove the following inequalities:
 
\begin{proposition}\label{teor.2C}
Assume that the differential operator~$\SL$ in~\eqref{SL} is regular at~$0$ and limit point at~$b$. Let~$H$ be the self-adjoint operator defined by $\SL$ satisfying the boundary conditions~\eqref{C.C.}. Given $E\in\R$ and $\varepsilon>0$, then
\begin{equation}\label{Des.JLC}\frac{5-\sqrt{24}}
{|m(E+i\varepsilon)|}<
\frac{\|u_{1,E}\|_{L(\varepsilon)}}
{\|u_{2,E}\|_{L(\varepsilon)}}<
\frac{5+\sqrt{24}}{|m(E+i\varepsilon)|}.
\end{equation}
\end{proposition}

Relation~\eqref{Des.JLC} allows for a generalization of Hausdorff subordinacy results  \cite{BCD, JL1,JL2, JZ} to obtain dimensional properties of the spectral measure of~$H$  in  bounded intervals. The next result follows directly by  Proposition~\ref{teor.2C}  and relation~\eqref{comp.l}.

\begin{corollary}\label{corolSL}
 Let~$H$ be as in Proposition~\ref{teor.2C},  $E\in\R$, $\varepsilon >0$ and a fixed $\kappa\in (0,1]$. Then 
\[
\limsup_{\varepsilon \to 0}\varepsilon^{(1-\kappa)}|m(E+i\varepsilon)|=\infty \ \Longleftrightarrow \ \liminf_{L \to b}
\frac{\|u_{1,E}\|_{L}}
{\|u_{2,E}\|^{\kappa /(2-\kappa)}_{L}}=0.
\]

\end{corollary}

 Given $\kappa\in (0,1]$, a solution~$u$ (for fixed~$E$) to~\eqref{EAC} is called $\kappa$-Hausdorff  subordinate at~$ b $ if
\[
\liminf_{L\rightarrow b}\frac{\|u\|_{L}}{\|v\|^{\kappa/(2-\kappa)}_{L}}=0,\]
for any solution~$v$  to~\eqref{EAC} linearly independent with~$u$. The (original) subordinate notion~\cite{G, GP} is recovered by taking~$\kappa=1$.

 Inspired by known results presented in Chapter~7 of~\cite{P}, we have

\begin{theorem}\label{teor.7.3.C} Let $\rho$ be the spectral  measure associated with the operator~$H$,  as in Proposition~\ref{teor.2C}, with boundary condition~\eqref{C.C.} at point 0 with $\varphi\in [0, 2\pi)$ fixed.  Pick~$\kappa$~$\in$~$(0,1]$ and let~$\mathcal I$ be a subset of the spectrum of~$H$ with $\rho( {\mathcal I})>0$.
\begin{enumerate}
	\item [(i)] Suppose that for all~$E\in {\mathcal I}$, the solution $u_{E}$ to~\eqref{EAC} is $\kappa$-Hausdorff  subordinate at the point~$ b$. Then, $\overline D_{\rho}^{\kappa}(E)=\infty$ for all~$E\in {\mathcal I}$, in particular the restriction $\rho( {\mathcal I}\cap\cdot)$ is $\kappa$-Hausdorff singular.
	\item [(ii)] Suppose that for all~$E\in {\mathcal I}$, there is no solution to~\eqref{EAC} that is  $\kappa$-Hausdorff  subordinate at~$b$. Then,  $\overline D_{\rho}^{\kappa}(E)<\infty$ for all~$E\in {\mathcal I}$, in particular the restriction $\rho( {\mathcal I}\cap\cdot)$ is $\kappa$-Hausdorff continuous. 
	\end{enumerate}
\end{theorem}

\begin{proof} 
By  general  results on Hausdorff measures and corresponding decompositions with respect to the behavior of  $\kappa$-derivative $\overline D_{\rho}^{\kappa}$ (see~\cite{ delRio, F, M, Ro}), this theorem is a simple consequence of  Corollary~\ref{corolSL} and the definition of $\kappa$-subordinate solution. 

More precisely, we have 
\begin{equation}\label{eq.derivada}
    \overline D_{\rho}^{\kappa}(E)=\infty\; \Longleftrightarrow\; \limsup_{\varepsilon \to 0}\varepsilon^{(1-\kappa)}|m(E+i\varepsilon)|=\infty.
\end{equation}

\begin{enumerate} \item [(i)] If for all~$E$, in some subset $ {\mathcal I}$, the solution $u_{E}$ to equation~\eqref{EAC} is $\kappa$-Hausdorff  subordinate at the point~$ b$, then, by Corollary~\ref{corolSL}, we have that
\[\limsup_{\varepsilon \to 
0}\varepsilon^{(1-\kappa)}|m(E+i\varepsilon)|=\infty \ .\]
Consequently, by equation~\eqref{eq.derivada} we have $\overline D_{\rho}^{\kappa}(E)=\infty,$ for all $E\in {\mathcal I}$. Thus  the restriction $\rho( {\mathcal I}\cap\cdot)$ is $\kappa$-Hausdorff singular.

\item [(ii)] If for all~$E$, in some subset $ {\mathcal I}$, there is no solution to equation~\eqref{EAC} that is $\kappa$-Hausdorff   subordinate at~$b$, then, by Corollary~\ref{corolSL}, 
\[\liminf_{\varepsilon \to 0}\varepsilon^{(1-\kappa)}|m(E+i\varepsilon)|<\infty.\]
Similarly, by equation~\eqref{eq.derivada}, we have $\overline D_{\rho}^{\kappa}(E)<\infty,$ for all $E\in {\mathcal I}$. Then, the restriction $\rho( {\mathcal I}\cap\cdot)$ is $\kappa$-Hausdorff continuous.
\end{enumerate}
\end{proof}

% SSSSS
\section{Asymptotic behavior of solutions}\label{solution}

In this section we use some notations presented in Appendix~\ref{M.S.C.}. Pick a decreasing sequence $(a_{n})$  of positive numbers and set $b_{n}=\sum_{j=1}^{n}8a_j$ so that 
\[
b=\lim_{n\rightarrow\infty}b_{n}=\sum_{n=1}^{\infty}8a_n<\infty,
\]
and consider the matrices in equations~\eqref{matrizaux} and~\eqref{matrizaux2}, respectively,   
\[
M_{n,n-1}(E)=\left(\begin{array}{cc} 1 -\frac{5}{3}E+O(a_n^{1/2}) & 1+O(a_n^{1/2}) \\ -\frac{5}{3}E+O(a_n^{1/2}) & 1+O(a_n)\end{array}\right),\quad M(E) = \left(\begin{array}{cc} 1 -\frac{5}{3}E & 1 \\ -\frac{5}{3}E & 1\end{array}\right).
\]

To simplify the notation, we will study the behavior of the solutions to the eigenvalue equation
\begin{equation}\label{mtsc}
-u''(x)+V(x)u(x)=Eu(x),
\end{equation}
with  potential~$V(x)$ of the form~\eqref{psc}, and will use  Pr\"ufer variables $R$ and~$\theta$ to analyze the behaviors of its solutions. First write
\begin{equation}\label{mv}
\left(\begin{array}{c} u_{E}(b_n) \\ u'_{E}(b_n)\end{array}\right)=p_nf_{+}+q_{n}f_{-},
\end{equation}
with eigenvectors
\[
f_{+}=\left(\begin{array}{c} 1 \\ 1-e^{-i\phi}\end{array}\right) \ \ \ 
\textrm{and} \ \ \ f_{-}=\left(\begin{array}{c} 1 \\ 
1-e^{i\phi}\end{array}\right)
\] 
 of  the matrix $M(E)$, associated with eigenvalues $e^{\pm i\phi}$, respectively. Define $R_{n}>0$ and $\theta_{n}\in\mathbb R$  by
\[p_{n}=iR_n e^{i\theta_n} \ \ \ \ \ \textrm{and} \ \ \ \ \ \ q_{n}=-iR_n e^{-i\theta_n},
\]
satisfying the initial conditions
\begin{align}\label{condicoes}
-2R_0\sin{\theta_0} = \cos{\varphi} \ \  \ \ \ \ \ \ \ \ \ \ \ \ \ \ \ \  -2R_0 \sin{\theta_0} +2R_0 \sin(\theta_0-\phi)  = \sin{\varphi} \ ,
\end{align}
where $\theta_0$ is chosen in $[0,2\pi)$. By relation~\eqref{norman}, we can write, for each $b_n>0$,
\[
\int_{0}^{b_{n}}[u_{E}(x)]^{2} \mathrm dx= u'_{E}(b_{n})\frac{\mathrm d}{\mathrm d E}u_{E}(b_{n})-u_{E}(b_{n})\frac{\mathrm d}{\mathrm d E}u'_{E}(b_{n}).\]
By~\eqref{mv},
\[
u_{E}(b_n)=-2R_n\sin \theta_n, \ \ \ \ \ \ \ \   u'_{E}(b_n)=2R_n[-\sin \theta_n + \sin (\theta_n -\phi)],
\] and so
\begin{equation}\label{eq.normabn}
\|u_E\|_{b_n}^2=\int_{0}^{b_{n}}[u_{E}(x)]^{2} \mathrm dx=4R_n^2\,\frac{\partial {\theta}_n}{\partial E}\,\sin\phi 
\end{equation} 
We use  transfer matrices to  analyze the behavior of $R_n$ and $\frac{\partial {\theta}_n}{\partial E}$ for large values of~$n$.

We have, by the construction of the potential~$V(x)$ and~\eqref{mv}, that
\begin{eqnarray*}
\left(\begin{array}{c} u_{E}(b_n) \\ u'_{E}(b_n)\end{array}\right)&=& \begin{pmatrix} 1 & 0 \\ g_n & 1 \end{pmatrix} M_{n,n-1}(E)\left(\begin{array}{c} u_{E}(b_{n-1}) \\ u'_{E}(b_{n-1})\end{array}\right)\\
&=& \begin{pmatrix} 1 & 0 \\ g_n & 1 \end{pmatrix} (M(E)+O(a_n))\left(\begin{array}{c} u_{E}(b_{n-1}) \\ u'_{E}(b_{n-1})\end{array}\right)\\
&=&\begin{pmatrix} 1 & 0 \\ g_n & 1 \end{pmatrix} ( p_{n-1}e^{i\phi} f_++ q_{n-1}e^{-i\phi}   f_-)+O(a_n) \\
&=& \begin{pmatrix} 1 & 0 \\ g_n & 1 \end{pmatrix} 
\begin{pmatrix}
p_{n-1}e^{i\phi}+q_{n-1} e^{-i\phi} \\
p_{n-1}e^{i\phi}(1-e^{-i\phi})+q_{n-1}e^{-i\phi}(1-e^{i\phi}) 
\end{pmatrix} +O(a_n) \\
&=& \begin{pmatrix}
p_{n-1}e^{i\phi} +q_{n-1}e^{-i\phi} \\
g_n (p_{n-1}e^{i\phi}+q_{n-1}e^{-i\phi}) +p_{n-1}(e^{i\phi}-1)+q_{n-1}(e^{-i\phi}-1)
   \end{pmatrix} +O(a_n),
\end{eqnarray*}
and again by~\eqref{mv}, 
\begin{equation}\label{mv2}
\left(\begin{array}{c} u_{E}(b_n) \\ u'_{E}(b_n)\end{array}\right)=\begin{pmatrix} p_n+q_n \\
               p_n(1-e^{-i\phi}) + q_n(1-e^{i\phi})
              \end{pmatrix} .
\end{equation}
Thus, we conclude that
\begin{align*}
p_n+q_n & = p_{n-1}e^{i\phi}+q_{n-1}e^{-i\phi} +O(a_n) \\
p_n(1-e^{-i\phi})+q_n(1-e^{i\phi}) &= g_n (p_{n-1}e^{i\phi}+q_{n-1}e^{-i\phi}) +p_{n-1}(e^{i\phi}-1) +q_{n-1}(e^{-i\phi}-1)+O(a_n).
\end{align*}

We need recurrence relations to~$p_n$ and~$q_n$. Multiplying the above first equation by $(1-e^{i\phi})$ and subtracting the second one, 
\begin{align*}
 p_n(-e^{i\phi}+e^{-i\phi}) & =
p_{n-1}e^{i\phi}(1-e^{i\phi})+q_{n-1}e^{-i\phi}(1-e^{i\phi}) -g_n(p_{n-1}e^{i\phi}+q_{n-1}e^{-i\phi})\\
&\;\;-p_{n-1}(e^{i\phi}-1) -q_{n-1}(e^{-i\phi}-1)+O(a_n) \\
&= p_{n-1}(1-e^{2i\phi}) -g_n(p_{n-1}e^{i\phi}+q_{n-1}e^{-i\phi})+O(a_n). 
\end{align*}
Therefore
\begin{align} \label{relacao_pn}
p_n &= p_{n-1} e^{i\phi} \left(1-\frac{i g_n}{2\sin{\phi}}\right)-\frac{i g_n e^{-i\phi}}{2\sin{\phi}} q_{n-1} +O(a_n)\ ,
\end{align}
\begin{align} \label{relacao_qn}
q_n &= q_{n-1} e^{-i\phi} \left(1+\frac{i g_{n}}{2\sin{\phi}}\right)+\frac{i g_n e^{i\phi}}{2\sin{\phi}} p_{n-1} + O(a_n) \ .
\end{align}
Relation~\eqref{relacao_qn} can be obtained with a calculation similar to the one done to obtain $p_n$ in terms of $(p_{n-1},q_{n-1})$, i.e., multiplying by $(1-e^{-i\phi})$, or equivalently using that $q_n=p_n^*$.

From $p_nq_n=R_n^2$, we have
\begin{equation} \label{relacaoR2}
\frac{R_n^2}{R_{n-1}^2}  =  1+\frac{g_n^2}{2\sin^2\phi} - \frac{g_n}{\sin{\phi}} \left( \sin{2(\theta_{n-1}+\phi)}+ \frac{g_n \cos{2(\theta_{n-1}+\phi)}}{2\sin{\phi}} \right)  +O(a_n). 
\end{equation}
To simplify the notation we denote $\tilde{\theta}_{n}={\theta}_{n-1}+\phi$ and equation~\eqref{relacaoR2} can be rewritten as
\begin{equation}\label{relacaoR}
\frac{R_{n}^{2}}{R_{n-1}^{2}}=1-\frac{g_n}{\sin\phi}\sin(2\tilde{\theta}_{n})+\frac{g_{n}^{2}}{\sin^{2}\phi}\sin^{2}(\tilde{\theta}_{n})+O(a_n).
\end{equation}
On the other hand, multiplying~\eqref{relacao_pn} by itself, we obtain 
\[
e^{2i{\theta}_{n}}=\frac{R_{n-1}^2}{R_n^2}\left[e^{2i\tilde{\theta}_{n}}-\eta_n^2(e^{2i\tilde{\theta}_{n}}-e^{-2i\tilde{\theta}_{n}}+2)-2i\eta_n(e^{2i\tilde{\theta}_{n}}+1)+O(a_n))\right],
\]
with  $\eta_n=\frac{g_n}{2\sin\phi}$. We can also write  relation~\eqref{relacaoR} as
\[
\frac{R_{n}^2}{R_{n-1}^2}= 1+\eta_n^2(2-e^{2i\tilde{\theta}_{n}}-e^{-2i\tilde{\theta}_{n}})+i\eta_n(e^{-2i\tilde{\theta}_{n}}-e^{2i\tilde{\theta}_{n}})+O(a_n).
\]
Thus,
\begin{equation}\label{relacaoTheta}
e^{2i{\theta}_{n}}=\frac{e^{2i\tilde{\theta}_{n}}+\eta_n^2(e^{2i\tilde{\theta}_{n}}-e^{-2i\tilde{\theta}_{n}}+2)-i\eta_n(e^{2i\tilde{\theta}_{n}}+1)+O(a_n)}{1+\eta_n^2(2-e^{2i\tilde{\theta}_{n}}-e^{-2i\tilde{\theta}_{n}})+i\eta_n(e^{-2i\tilde{\theta}_{n}}-e^{2i\tilde{\theta}_{n}})+O(a_n)}.
\end{equation}

\begin{proposition} \label{limiteTheta}
There is a parameter~$ C>0$, depending on $\phi  \in (0,\pi)$, such that
\[
\lim_{n\rightarrow \infty}\frac{1}{n}\left|\frac{\partial {\theta}_n}{\partial E}\right| = C,
\]
with $\cos\phi=1-\frac{5}{6}E$.
\end{proposition}

\begin{proof}
To prove this bound on $\frac{\partial {\theta}_n}{\partial E}$, it suffices to consider  $\frac{\partial {\theta}_n}{\partial \phi}$, because $\frac{\partial {\theta}_n}{\partial E}=\frac{5}{6\sin \phi}\frac{\partial {\theta}_n}{\partial \phi}$. Since $\lim g_n=0$  and remembering that  $\eta_n=\frac{g_n}{2\sin\phi}$, we can write~\eqref{relacaoTheta}  as
\begin{equation*}
e^{2i{\theta}_{n}}=\frac{e^{2i\tilde{\theta}_{n}}}{1+z(\theta_n,\phi)}  
+t(\theta_n,\phi) ,
\end{equation*}
with 
\[
z(\theta_n,\phi)=\eta_n^2(2-e^{2i\tilde{\theta}_{n}}-e^{-2i\tilde{\theta}_{n}})+i\eta_n(e^{-2i\tilde{\theta}_{n}}-e^{2i\tilde{\theta}_{n}})+O(a_n)
\] and 
\[
t(\theta_n,\phi)=\frac{\eta_n^2(e^{2i\tilde{\theta}_{n}}-e^{-2i\tilde{\theta}_{n}}+2)-i\eta_n(e^{2i\tilde{\theta}_{n}}+1)+O(a_n)}{1+\eta_n^2(2-e^{2i\tilde{\theta}_{n}}-e^{-2i\tilde{\theta}_{n}})+i\eta_n(e^{-2i\tilde{\theta}_{n}}-e^{2i\tilde{\theta}_{n}})+O(a_n)},
\] which are continuously differentiable functions and converge uniformly to zero as $n\rightarrow \infty$; hence, we obtain the estimate
\begin{equation}\label{limiteThetaeq}
    \left|\frac{\partial {\theta}_n}{\partial \phi}\right|\leq C_n\left(\left|\frac{\partial {\theta}_{n-1}}{\partial \phi}\right|+1\right) + \varepsilon_n,
\end{equation}
with the sequences $C_n \rightarrow 1$ and $\varepsilon_n \rightarrow 0$. 

Assume inductively that $\frac{\partial {\theta}_{n-1}}{\partial 
\phi}=O(n-1)$, that is,  there is a constant $D$ such that 
\[
\left|\frac{\partial \tilde{\theta}_{n-1}}{\partial \phi}\right|\leq D(n-1),
\]
and observe that the induction base case follows from the initial 
conditions~\eqref{condicoes}. Then, by relation \eqref{limiteThetaeq}, 
\[
 \left|\frac{\partial {\theta}_n}{\partial \phi}\right|\leq n\left(DC_n+\frac{(1-D)}{n}C_n + \frac{\varepsilon_n}{n}\right) .
\]
Since $C_n \rightarrow 1$ and $\varepsilon_n \rightarrow 0$ as $n\to \infty$, 
we conclude that $ \frac{\partial {\theta}_n}{\partial \phi}= O(n)$. Therefore, 
there is a  $ C>0 $, depending on $\phi \in (0,\pi)$, so that
\[
\lim_{n\rightarrow \infty}\frac{1}{n}\left|\frac{\partial {\theta}_n}{\partial E}\right| = C.
\]
This completes the proof of the proposition.
\end{proof}

Note that, by equation~\eqref{eq.normabn}, there is a $0<C<\infty$ so that
\begin{equation}\label{eq.normabn2}
\lim_{n\rightarrow \infty}\frac{\|u_E\|^2_{b_{n}}}{n}=\lim_{n\rightarrow \infty}\frac{1}{n}\int_{0}^{b_{n}}[u_{E}(x)]^{2}\mathrm dx = C \lim_{n\rightarrow \infty}R_n^2.
\end{equation}

We conclude that it is possible to obtain information about the asymptotic 
behavior of the solutions~$u_{E}$ by studying the behavior of the sequence 
$(R_{n})$, as $n\rightarrow\infty$.  In the next section we  analyze the sequence $(R_{n})$ for  selected $(g_n)$, so concluding the proof of Theorem~\ref{thmSubord}.

%  SSSSS
\section{Proof  of Theorem~\ref{thmSubord}}\label{sectProofThmSub}

 We note that the estimates for the Pr\"ufer variables in \eqref{relacaoR} coincide, up to terms of $O(a_n)$, with the estimates in \cite{KLS2} (Equation (2.12.c)) for discrete Schr\"odinger operators. Since the sequence $(a_n)$ tends to zero, these terms are not expected  to interfere with the asymptotic behavior of  the sequence~$(R_{n})$.  We will justify this expectation in the following.

We specialize to the potential~$V_{\omega}(x)$  of the form~\eqref{psc} with independent random variables $g_{n}\equiv g_n^{\omega}$ defined in a probability space,  with (probability) measure~$\nu(\omega)$, and satisfying the conditions (i)-(iv)  in Subsection~\ref{subsectPot}.

\begin{proposition}\label{versaoTeo8.2}
Suppose that the sequence $(g_{n}^{\omega})$ satisfy  (i)-(iv) in Subsection~\ref{subsectPot} . 
Fix~$\phi\in(0,\pi)$, with $\phi\neq 
\frac{\pi}{4},\frac{\pi}{2},\frac{3\pi}{4}$. 
Then,  for $\nu$-a.e.~$\omega$, the variables $R_n$, associated with  a solution to the eigenvalue equation~\eqref{mtsc}, satisfy
\begin{equation}\label{betapositivo}
\lim_{n\rightarrow \infty}\frac{\ln R_{n}}{\left(\sum_{j=1}^{n}\frac{1}{j}\right)}=\frac{\lambda^{2}}{8\sin^{2}\phi}.
\end{equation}
\end{proposition}
\begin{proof}

Fix  $E$ and an initial condition $\theta_{0}$ associated with a solution to~\eqref{mtsc}; by~\eqref{relacaoR},
\[
\ln R_{n}-\ln R_{n-1}=\frac{1}{2}\ln\left(1-\frac{g_n^{\omega}}{\sin\phi}\sin(2\tilde{\theta}_{n})+\frac{(g_{n}^{\omega})^{2}}{\sin^{2}\phi}\sin^{2}(\tilde{\theta}_{n})+y_n\right),\]
with $(y_{n})$ a sequence so that $|y_{n}|\leq C a_{n}$, for a~$C>0$ 
independent of~$n$. Since
\[
\sup_{\omega}\left|\frac{g_n^{\omega}}{\sin\phi}\sin(2\tilde{\theta}_{n})+\frac{(g_{n}^{\omega})^{2}}{\sin^{2}\phi}\sin^{2}(\tilde{\theta}_{n})+y_n\right|\longrightarrow 0,
\]
as $n\rightarrow \infty$, we may use the expansion 
\[
\ln(1+x)=x-\frac{x^{2}}{2}+O(x^{3})
\]
to get 
\begin{eqnarray*}
\ln\Big(1-\frac{g_n^{\omega}}{\sin\phi}\sin(2\tilde{\theta}_{n}) &+& 
\frac{(g_{n}^{\omega})^{2}}{\sin^{2}\phi}\sin^{2}(\tilde{\theta}_{n})+y_n\Big) 
\\ 
&=& -\frac{g_n^{\omega}}{\sin\phi}\sin(2\tilde{\theta}_{n})+
\frac{(g_{n}^{\omega})^{2}}{\sin^{2}{\phi}}\sin^{2}(\tilde{\theta}_{n})\\
 & & + \ y_n -\frac{(g_{n}^{\omega})^{2}}{2\sin^{2}\phi}\sin^{2}(2\tilde{\theta}_{n})\\
& &  + \ O\left((g_{n}^{\omega})^{3}+a_n\right).
\end{eqnarray*}
Now, by using the trigonometric relation
\begin{eqnarray*}\sin^{2}\theta-\frac{1}{2}\sin^{2}(2\theta)&=&\frac{1}{2}-\frac{1}{2}\cos(2\theta) -\frac{1}{2}\left[\frac{1}{2}-\frac{1}{2}\cos(4\theta)\right]\\
&=&\frac{1}{4}-\frac{1}{2}\cos(2\theta)+\frac{1}{4}\cos(4\theta),\end{eqnarray*}
we obtain that
\[\ln R_{n}=\frac{1}{8}\sum_{n=1}^{\infty}\frac{\E((g_{n}^{\omega})^{2})}{\sin^{2}\phi}+ C_{1} + C_2 + C_3 + C_4,\]
with corrections
\begin{equation*}
\left. \begin{array}{l} C_1 = -\frac{1}{2\sin\phi}\sum_{j=1}^{n}  g_{j}^{\omega}\sin(2\tilde{\theta}_{j})\\
C_2 = \frac{1}{2\sin^{2}\phi}\sum_{j=1}^{n} \left[((g_{j}^{\omega})^{2}) - \E((g_{j}^{\omega})^{2})\right] \left[\sin^{2}(\tilde{\theta}_{j})-\frac{1}{2}\sin^{2}(2\tilde{\theta}_{j})\right]\\
C_3 = \frac{1}{2\sin^{2}\phi}\sum_{j=1}^{n}  (\E(g_{j}^{\omega})^{2}) \left[\frac{1}{2}\cos(2\tilde{\theta}_{j})-\frac{1}{4}\cos(4\tilde{\theta}_{j})\right]\\
C_4 = \sum_{j=1}^{n} \mathrm{O}\left((g_{j}^{\omega})^{3}+a_{j}\right).
\end{array}\right.
\end{equation*}

Hence, the result  follows if we prove that for each $q=1,2,3,4$ and a.e.~$\omega$,
\[\lim_{n\rightarrow \infty} \frac{|C_{q}(\omega)|}{\left(\sum^{n}_{j=1}\frac{1}{j}\right)}=0.\]
 For $q=1,2,3$, the above limit follows exactly as in Theorem~$8.2$ in 
\cite{KLS2}, since we have analogous expressions for $C_1,C_2,C_3$; here it is necessary that $\phi\neq 
\frac{\pi}{4},\frac{\pi}{2},\frac{3\pi}{4}$.

For $q=4$, the result follows from  hypothesis (iii) and the construction of
the sequence~$(a_n)$, since $\sum_{n=1}^{\infty}8a_n=b<\infty$.
\end{proof}

\begin{proposition}\label{versaoLema8.8}
Let $(g_{n}^{\omega})$ be as in Proposition~\ref{versaoTeo8.2}. Fix 
$\phi\in(0,\pi)$, with $\phi\neq \frac{\pi}{4},\frac{\pi}{2},\frac{3\pi}{4}$. 
Then, for $\nu$-a.e.~$\omega$, there is a solution~$u_{\theta_{\omega}}$ to~\eqref{mtsc} so that $R^{(\theta_{\omega})}$ satisfies
\begin{equation}\label{betanegativo}
\lim_{n\rightarrow \infty}\frac{\ln R^{(\theta_{\omega})}_{n}}{\ln n}=-\frac{\lambda^{2}}{8\sin^{2}\phi}.
\end{equation}
\end{proposition}

\begin{proof}
Let $\beta=\frac{\lambda^{2}}{8\sin^{2}\phi}$. Let $R^{(1)}_{n}$ and $R^{(2)}_{n}$ be the radial Pr\"ufer variables associated with $\theta_{n}^{(1)}$ and $\theta_{n}^{(2)}$, respectively. By Proposition~\ref{versaoTeo8.2}, for a.e.~$\omega$ and $k=1,2$,
\begin{equation}\label{limRbeta}
\lim_{n\rightarrow \infty}\frac{\ln R_{n}^{(k)}}{\ln n}=\beta.\end{equation}
By  relation~\eqref{mv2}, we have that
 \[u_{E}(b_n)=-2R_n \sin({\theta}_{n}),\] \[u'_{E}(b_n)=-2R_n[\sin(\phi)\cos({\theta}_{n})+(1-\cos\phi)\sin({\theta}_n)].\] 
Then, for any two linearly independent solutions $u_{1,E}$ and $u_{2,E}$ to  equation~\eqref{mtsc},  associated with $R^{(k)}$ and $\theta^{(k)}$, $k=1,2$, respectively, we have 
\begin{eqnarray*}
W[u_{1,E},u_{2,E}](b_{n})&=&u_{1,E}(b_n)u'_{2,E}(b_n)-u'_{1,E}(b_n)u_{2,E}(b_n)\\
&=&4R_{n}^{(1)}R_{n}^{(2)}\sin\phi\left(\sin({\theta}^{(1)}_n)\cos({\theta}_n^{(2)})-\sin({\theta}_n^{(2)})\cos({\theta}_n^{(2)}))\right)\\
&=&4R_{n}^{(1)}R_{n}^{(2)}\sin\phi\sin(\theta_{n}^{(1)}-\theta_{n}^{(2)}).
\end{eqnarray*}

Since the Wronskian is constant, one gets 
\[
R_{n}^{(1)}R_{n}^{(2)}\sin\phi\sin(\theta_{n}^{(1)}-\theta_{n}^{(2)})=C_{\alpha}.
\] So, by~\eqref{limRbeta}, 
\[\lim_{n\rightarrow\infty}\frac{\ln\big|\sin\big(\theta_{n}^{(1)}-\theta_{n}^{(2)}\big)\big|}{\ln n}=-2\beta.
\]
Therefore, to conclude this proof it is enough to follow the same steps of the proof of  Lemma~$8.8$ in~\cite{KLS2} (from equation~(8.20) onwards).
\end{proof}

Our Propositions~\ref{versaoTeo8.2} and~\ref{versaoLema8.8} are versions, in our setting, of Theorem~$8.2$ and Lemma~$8.8$ in~\cite{KLS2}, respectively. These theorems are for~$\phi\neq\frac{\pi}{4},\frac{\pi}{2},\frac{3\pi}{4}$.

Recalling that in the model discussed here we have~$\cos\phi=1-\frac{5}{6}E$, it follows that $E\in(0,\frac{12}{5})$ with
\[
E\neq \frac{3(2-\sqrt{2})}{5},\,\frac{6}{5},\, \frac{3(2+\sqrt{2})}{5}.
\] 
By Proposition~\ref{versaoLema8.8} and  relation~\eqref{eq.normabn}, it 
follows that for $\nu$-a.e.~$\omega$ there is a solution $u_{E}^{S}$ to 
equation~\eqref{mtsc} so that (the superscript~$S$ is for ``subordinate;'' see Subsection~\ref{subsectDemoSingCont})
\begin{equation}\label{equivalencia1}
\|u_{E}^{S}\|_{b_{n}} \asymp n^{-\beta}n^{1/2},
\end{equation}
with $r_1(n)\asymp r_2 (n)$ denoting the relation $\lim_{n\to\infty}\frac{\ln r_1}{\ln r_2}=C$, for some $0<C<\infty$.
Note that if $\beta<1/2$, then $\lim_{n\rightarrow\infty}\|u_{E}^{S}\|_{b_{n}}=\infty,$ and consequently, $u_{E}^{S}\notin \mathrm{L}^{2}(I_b)$.

Similarly, by Proposition~\ref{versaoTeo8.2} and  relation~\eqref{eq.normabn}, it follows that  for every solution~$u_{E}$ to equation~\eqref{mtsc}, with $u_{E}$ linearly independent with~$u_{E}^S$, we have
\begin{equation}\label{equivalencia2}\|u_{E}\|_{b_{n}} \asymp n^{\beta}n^{1/2}.\end{equation}

\begin{lemma} \label{sequenciaL}
Suppose $\|u\|_{b_n} \asymp n^{\beta+1/2}$. If $L_j$ is a sequence in $(0,b)$ with $L_j\uparrow b$, for large~$j$ pick the (unique) subsequence $(b_{n_j})$ so that $b_{n_j}\le L_j<b_{n_j+1}$. Then $\|u\|_{L_j} \asymp n_{j}^{\beta+1/2}$.
\end{lemma}

\begin{proof}
Since $\|u\|_L$ is a monotone function of $L$, one has
\[ \|u\|_{b_{n_j}}\leq \|u\|_{L_j} < \|u\|_{b_{n_j+1}} 
\] and so
\[
\frac{\ln \|u\|_{b_{n_j}}}{\ln n_j^{\beta+1/2}} \;\leq \;\frac{\ln\|u\|_{L_j}}{\ln n_j^{\beta+1/2}} \; <\; \frac{\ln \|u\|_{b_{n_j+1}}}{\ln (n_j+1)^{\beta+1/2}}\times \frac{\ln (n_j+1)}{\ln n_j}
\] and since $\ln (n_j+1)/\ln n_j\downarrow 1$ as $j\to\infty$, the result follows.
\end{proof}

\begin{proof}(\textit{Theorem~\ref{thmSubord}})

Fix $0<\lambda<2$ and consider the Schr\"odinger operator generated by a potential of the form~\eqref{psc} satisfying (i)-(iv). Let $E\in J(\lambda)$, with~$J(\lambda)$ as in Subsection~\ref{subsecMainRes}. 

From $\cos\phi=1-\frac{5}{6}E$, for  $\phi\in(0,\pi)$, we have \[\sin^{2}\phi=1-\cos^{2}(\phi)=\frac{60E-25E^{2}}{36},\]
and since $\beta=\frac{\lambda^{2}}{8\sin^{2}\phi}$, one finds
 \[\beta=\frac{9\lambda^{2}}{120E-50E^{2}}.\]
Thus $\beta<\frac{1}{2}$ if and only if $\lambda<2$ and
$E\in J$. We have, by~\eqref{equivalencia1} and~\eqref{equivalencia2}, that  $
\|u_{E}^{S}\|_{b_{n}}  \asymp n^{\frac{1}{2}-\beta}
$, and for $u_E$ linearly independent with $u_{E}^{S}$, one has  
$
\|u_{E}\|_{b_{n}} \asymp n^{\frac{1}{2}+\beta}.
$

Noting that $\left(\frac{1}{2}-\beta\right) -\tilde{\alpha}\left(\frac{1}{2}+\beta\right)=0$ if and only if   $\tilde{\alpha}=\frac{1-2\beta}{1+2\beta}$, consequently 
\[
\lim_{n\rightarrow\infty}\frac{\|u_{E}^{S}\|_{b_{n}}}{\|u_{E}\|^{\tilde{\alpha}}_{b_{n}}}=A,\
\]
for some $0<A<\infty$. Hence, we conclude, by Lemma~\ref{sequenciaL} that for $0<L<b$,
\begin{equation}\label{eqfianl1}
\lim_{L\rightarrow b}\frac{\|u_{E}^{S}\|_L}{\|u_{E}\|^{\alpha/(2-\alpha)}_{L}}=A
\end{equation}
with $\tilde{\alpha}=\frac{\alpha}{2-\alpha}$, and any solution~$u_{E}$ to~\eqref{mtsc}, linearly independent  with~$u_{E}^{S}$.
\end{proof}

\subsection{Singular continuous spectra}\label{subsectDemoSingCont}
The fact that the Schr\"odinger operator, with a potential of the form~\eqref{psc} and $\sum_n g_n^2=\infty$, has singular continuous spectrum in $(0,12/5)$,  follows by Proposition~3 of~\cite{P1978} (since there are similarities in the calculations in our Section~\ref{solution} with those in (\cite{P1978},  pages 32--35)). However,~\cite{P1978} does not provide information on  Hausdorff dimensions.

Another way to verify that the operator~\eqref{eqHomega}, with $0<\lambda<2$ and a potential of the form~\eqref{psc}  satisfying (i)-(iv), has singular continuous spectrum in~$J(\lambda)$ for $\nu$-a.s.~$\omega$,  is by noting that for $0<\beta<1/2$, then by relations~\eqref{equivalencia1}-\eqref{equivalencia2} and Lemma~\ref{sequenciaL}, one has
\[
\lim_{L\rightarrow\ b}\frac{\|u_{E}^{S}\|_{L}}{\|u_{E}\|_{L}}=0,
\] 
for any solution~$u_{E}$ to~\eqref{mtsc} linearly independent  with~$u_E^S$. Thus, for  all $E\in J$, it follows that $u_{E}^{S}$ is  subordinate at~$b$ (i.e., with~$\kappa=1$ in Section~\ref{SeI}), and  by~\eqref{equivalencia1}, one has $u_{E}^{S}\notin \mathrm{L}^{2}[0,b]$ and $E$ is not an eigenvalue; hence, by  Theorem~7.3 in~\cite{P},  the operator has a purely singular continuous spectrum in this set.  

\appendix 
\section{The potential $V_0$}\label{M.S.C.}

We recall the example from~\cite{P1975} of a family   Schr\"odinger operators $H=-\mathrm d^{2}/\mathrm d x^{2}+V_0(x)$, in the bounded interval~$I_b$, which has  purely absolutely continuous spectrum in certain interval of energies. 
For  $0<a<1$, let $V_{a}(x)$ be given by
\begin{eqnarray}\label{1potencial}
V_{a}(x)&=&y_1\delta(x-a)+y_2\delta(x-2a)+y_2\delta(x-3a)+y_1\delta(x-4a) \nonumber\\ 
& & +y_2\delta(x-5a)+y_1\delta(x-6a)+y_1\delta(x-7a)+y_2\delta(x-8a),\end{eqnarray}
with $y_1\equiv y_1(a)=(a^{-3/2}-a^{-1})$, $y_2\equiv y_2(a)=(a^{-1/2}-a^{-1})$; $\delta$ represents the Dirac~$\delta$ distribution.

Pick $f(x)$ a solution to the eigenvalue equation
\[-f''(x)+V_a(x)f(x)=Ef(x), \ \ \ \ (E>0).\] 
Since $f'(x)$ is discontinuous at each $\delta$-singularity $x=Ka$, with $K=1,\ldots,8$, we adopt the convention that $f'(Ka)$ is continuous from the right, that is, 
\[
f'(Ka)=\lim_{x\rightarrow Ka^+}f'(x).
\]
One then associates transfer matrices to such  eigenvalue equation as
\[\left(\begin{array}{c}f(a) \\ f'(a)\end{array}\right)=M^{(1)}_{a}\left(\begin{array}{c}f(0) \\ f'(0)\end{array}\right) \qquad \textrm{and} \qquad \left(\begin{array}{c}f(2a) \\ f'(2a)\end{array}\right)=M^{(2)}_{a}\left(\begin{array}{c}f(a) \\ f'(a)\end{array}\right),\]
with
\begin{eqnarray*}
M^{(1)}_{a}&=&\left(\begin{array}{cc} 1 & 0 \\ y_1&1\end{array}\right)\left(\begin{array}{cc} \cos(E^{1/2}a) & E^{-1/2}\sin(E^{1/2}a) \\ -E^{1/2}\sin(E^{1/2}a)& \cos(E^{1/2}a)\end{array}\right)\\
&=&\left(\begin{array}{cc} \cos(E^{1/2}a) & E^{-1/2}\sin(E^{1/2}a) \\ y_{1}\cos(E^{1/2}a)-E^{1/2}\sin(E^{1/2}a) & y_{1}E^{-1/2}\sin(E^{1/2}a)+\cos(E^{1/2}a)\end{array}\right),
\end{eqnarray*}
\begin{eqnarray*}
M^{(2)}_{a}&=&\left(\begin{array}{cc} 1 & 0 \\ y_2&1\end{array}\right)\left(\begin{array}{cc} \cos(E^{1/2}a) & E^{-1/2}\sin(E^{1/2}a) \\ -E^{1/2}\sin(E^{1/2}a)& \cos(E^{1/2}a)\end{array}\right)\\
&=&\left(\begin{array}{cc} \cos(E^{1/2}a) & E^{-1/2}\sin(E^{1/2}a) \\ y_{2}\cos(E^{1/2}a)-E^{1/2}\sin(E^{1/2}a) & y_{2}E^{-1/2}\sin(E^{1/2}a)+\cos(E^{1/2}a)\end{array}\right).
\end{eqnarray*}
Since we are  interested in obtaining estimates of the behavior of these matrices as $a\rightarrow 0$, by using  series expansions of sine and cosine functions,  one finds
\[M^{(1)}_{a}=\left(\begin{array}{cc} 1 +O(a^{2}) & a+O(a^{3}) \\ a^{-3/2}-a^{-1}-\frac{1}{2}Ea^{1/2}+O(a) & a^{-1/2}-\frac{1}{6}Ea^{3/2}+O(a^{2})\end{array}\right),\]
\[M^{(2)}_{a}=\left(\begin{array}{cc} 1-\frac{1}{2}Ea^{2} +O(a^{5/2}) & a+O(a^{3}) \\ a^{-1/2}-a^{-1}-\frac{1}{2}Ea+O(a^{3/2}) & a^{1/2}-\frac{1}{3}Ea^{2}+O(a^{5/2})\end{array}\right).\]  

The multiplication $M^{(2)}_{a}M^{(1)}_{a}$ gives the transfer matrix from position~$0$ to~$2a$. Similarly, the transfer matrix from~$0$ to~$8a$ is
\begin{equation}\label{matrizaux}
M_{a}(E)=M^{(2)}_{a}(M^{(1)}_{a})^{2}M^{(2)}_{a} M^{(1)}_{a}(M^{(2)}_{a})^{2}M^{(1)}_{a}=\left(\begin{array}{cc} 1 -\frac{5}{3}E+O(a^{1/2}) & 1+O(a^{1/2}) \\ -\frac{5}{3}E+O(a^{1/2}) & 1+O(a)\end{array}\right),
\end{equation} that is,
\[
\left(\begin{array}{c}f(8a) \\ f'(8a)\end{array}\right)=M_{a}(E)\left(\begin{array}{c}f(0) \\ f'(0)\end{array}\right).
 \]

We are now in  position to describe the potential $V_0(x)$ in $[0,b]$, defined 
in~\cite{P1975}, so that the operator $H=-\mathrm d^{2}/\mathrm d 
x^{2}+V_0(x)$ has  purely absolutely continuous spectrum in the interval 
$(0,12/5)$. Let $(b_n)$ be an increasing sequence in $[0,b)$ to be specified; in each 
sub-interval $[b_n,b_{n+1})$, the potential $V_0(x)$ is defined from 
$V_{a_{n}}(x)$ as
\begin{eqnarray}\label{potenciala.c}
V_0(x)&=& V_{a_1}(x)+V_{a_2}(x-8a_1)+V_{a_3}(x-8(a_1+a_2))+\nonumber\\
& & +V_{a_4}(x-8(a_1+a_2+a_3))+\ldots,
\end{eqnarray}
with $(a_{n})$ a decreasing sequence of positive numbers, $b_{n}=\sum_{j=1}^{n}8a_j$ so that 
\[
b=\lim_{n\rightarrow\infty}b_{n}=\sum_{n=1}^{\infty}8a_n<\infty.
\]
\begin{remark}
In the limit as $x\rightarrow b$, the amplitude of oscillations of~$V_0(x)$ 
tends to infinity, while the period of oscillation tends to zero; thus 
$V_0(x)$ is unbounded from both  above and below near~$x=b$.
\end{remark}
Let $\psi_{E}(x)$ be a solution to the equation
\begin{equation}\label{EAAC}
-\psi''_E(x)+V_0(x)\psi(x)=E\psi_E(x),
\end{equation} and  consider the transfer matrices  $M_{n}(E)$ and  $M_{m,n}(E)$ 
\[
\left(\begin{array}{c} \psi_{E}(b_n) \\ 
\psi'_{E}(b_n)\end{array}\right)=M_{n}(E)\left(\begin{array}{c} \psi_{E}(0) \\ 
\psi'_{E}(0)\end{array}\right),
 \]
\begin{equation}\label{MT1}
\left(\begin{array}{c} \psi_{E}(b_n) \\ \psi'_{E}(b_n)\end{array}\right)=M_{m,n}(E)\left(\begin{array}{c} \psi_{E}(b_m) \\ \psi'_{E}(b_m)\end{array}\right).
\end{equation}
By~\eqref{MT1} and the expression of~$V_0(x)$, we have  
\[M_{n,n+1}(E)=M_{a_n}(E),\]
with $M_{a_n}(E)$ as in~\eqref{matrizaux} for $a=a_{n}$. Since $\lim_{n\rightarrow\infty}a_{n}=0$, it follows that
\[\lim_{n\rightarrow\infty}\|M_{n,n+1}(E)-M(E)\|=0,\]
with
\begin{equation}\label{matrizaux2}
M(E)=\left(\begin{array}{cc} 1 -\frac{5}{3}E & 1 \\ -\frac{5}{3}E & 1\end{array}\right).\end{equation}

It is proven in \cite{P1975} that the above convergence is uniform for the values of~$ E $ in each closed subinterval of $(0,12/5)$. The choice of  $E\in (0,12 / 5) $  is  important, because within this interval $M(E)$ is elliptic with  distinct  complex eigenvalues  $e^{\pm i\phi}$, with $\cos \phi=1-5E/6$.

By Lemma~$3$ in~\cite{P1975}, we have
\begin{equation}\label{norman} 
\|\psi_{E}\|_{b_{n}}^2=\int_{0}^{b_{n}}[\psi_{E}(x)]^{2} \mathrm dx= \psi'_{E}(b_{n})\frac{\mathrm d}{\mathrm d E}\psi_{E}(b_{n})-\psi_{E}(b_{n})\frac{\mathrm d}{\mathrm d E}\psi'_{E}(b_{n}).\end{equation}
Hence,  $\|\psi_{E}\|_{b_{n}}$ can be analyzed through the behavior of~$\psi$ and~$\psi'$ at~$b_n$, which can be estimated by powers of the matrix~$M(E)$.  Such arguments  were employed in the proof of Lemma~$3$ in~\cite{P1975}, resulting in
\begin{equation}\label{subordcontinua}
0<\lim_{n\rightarrow\infty}\frac{\|\psi_{E}\|_{b_{n}}^{2}}{n}=\lim_{n\rightarrow\infty}\frac{1}{n}\int_{0}^{b_{n}}[\psi_{E}(x)]^{2} \mathrm dx <\infty.
\end{equation}

By using the relations recalled in this appendix,  in~\cite{P1975, P} it is proven that $-\mathrm d^2/\mathrm d x^{2}+V_0(x)$, with the boundary conditions~\eqref{C.C.}, has purely absolutely continuous spectrum in the  interval $(0, 12 / 5) $.

\begin{remark}
By relation~\eqref{subordcontinua}, any solutions $\psi_E$ in the interval of energy $(0,12/5)$ behaves asymptotically as $n^{1/2}$. The developments in Section~\ref{solution} were pursued with this given.
\end{remark}

\
 
 \noindent {\bf Acknowledgment.} CRdO thanks the partial support by Conselho Nacional de Desenvolvimento Cient{\'{\i}}fico e Tecnol\'ogico (under contract number 303503/2018-1).

\

\end{document}